# Title: On Energization and Loss of the Ionized Heavy Atom and Molecule in Mars' Atmosphere

**Authors:** J. -T. Zhao[1#], Q. -G. Zong[2,1*], Z. -Y. Liu[3], X. -Z. Zhou[1], S. Wang[1], W. -H. Ip[4], C. Yue[1], J. -H. Li[1], Y. -X. Hao[5], R. Rankin[6], A. Degeling[7], S. -Y. Fu[1], H. Zou[1], and Y. -F. Wang[1]

**Affiliations:**

[1]Institute of Space Physics and Applied Technology, Peking University, Beijing, China.

[2]Key Laboratory of Lunar and Planetary Sciences, Macau University of Science and Technology (MUST), Macau, China.

[3]Institut de Recherche en Astrophysique et Planétologie, CNES-CNRS-Universite Toulouse III Paul Sabatier, Toulouse, France.

[4]Institute of Astronomy, National Central University, Jhongli, Taiwan.

[5]Max Planck Institute for Solar System Research, Göttingen, Germany.

[6]Department of Physics, University of Alberta, Edmonton, AB, Canada.

[7]Shandong Provincial Key Laboratory of Optical Astronomy and Solar-Terrestrial Environment, Institute of Space Sciences, Shandong University, Weihai, China.

*Corresponding author. Email: qgzong@pku.edu.cn

#The first author, Jiutong Zhao, finished this manuscript in Peking University during his PhD era and has joined Space Sciences Laboratory - University of California, Berkeley since this August after the submission of this paper.

**Abstract:**

The absence of global magnetic fields is often cited to explain why Mars lacks a dense atmosphere. This line of thought is based on a prevailing theory that magnetic fields can shield the atmosphere from solar wind erosion. However, we present observations here to demonstrate a counterintuitive understanding: unlike the global intrinsic magnetic field, the remnant crustal magnetic fields can enhance atmosphere loss when considering loss induced by plasma wave-particle interactions. An analysis of MAVEN data, combined with observation-based simulations, reveals that the bulk of $O^+$ ions would be in resonance with ultra-low frequency (ULF) waves when the latter were present. This interaction then results in significant particle energization, thus enhancing ion escaping. A more detailed analysis attributes the occurrence of the resonance to the presence of Mars' crustal magnetic fields, which cause the majority of nearby ions to gyrate at a frequency matching the resonant condition ($\omega - k_{\parallel} v_{\parallel} = \Omega_i$) of the waves. The ULF waves, fundamental drivers of this entire process, are excited and propelled by the upstream solar wind. Consequently, our findings offer a plausible explanation for the mysterious changes in Mars' climate, suggesting that the ancient solar wind imparted substantially more energy.

**Main Text:**

Mars is believed to have undergone significant atmospheric loss during its transition from a warm and wet Noachian era (4.1 Giga-annum, $10^9$ years, Ga – 3.7 Ga) to the current cold and dry Amazonia era[1]. Understanding how this loss occurred is at the central of Mars research. When examining the possibility of the escape of heavy atom and molecular (e.g., O, $O_2$, $CO_2$), Jeans escape theory is believed to contribute very little because the energy (~2.0-5.5 eV) necessary to surpass gravitational bounds is much larger than the atmospheric temperature (~0.02 eV). In other words, additional energization is required to account for these heavy ions' escape.

Depending on the Martian plasma environment, various energization mechanisms could happen. Above the dayside magnetic pile-up boundary, the solar wind motional electric field prevails, and the process can be interpreted in terms of pick-up or mass-loading[2]. Within the magnetotail plasma sheet, the $\mathbf{j} \times \mathbf{B}$ force can accelerate and carry away the ionospheric ions[2,3]. Though the build-up of the electric field in the region beneath during the solar wind-Mars interaction is still in dispute, it determines the number of particles that can be transported to the solar wind dominant region and limits the total escape rate[4]. One widely-acknowledged energy source is the polarization electric field. However, this explanation encountered a bottleneck when interpreting the escape of molecular ions.

In addition to the stationary electric field, solar wind-induced ultra-low frequency waves can penetrate the upper ionosphere and implement energy delivery[5]. Fig. 1**A** illustrates the global scenario of the resonant energy delivery from the solar wind-induced ULF waves to the planetary ions. Periodic foreshock activities near proton gyro-frequency (~0.03 Hz for |B| = 2 nT) serve as a disturbance source, giving rise to ULF wave excitation within the ionosphere. This wave activity can permeate the ionosphere, reaching altitudes below approximately 400 km and persisting for around 10 minutes[6,7]. The resonant interaction between the ULF waves and ionospheric ions can result in ion energization and escape, which have been unveiled in numerical modeling[8,9]. Subsequently, ULF waves emerge as a promising conduit for transmitting solar wind energy.

It is noted that resonant energization is more likely to occur above regions with crustal remnant magnetic fields. In these areas, conditions conducive to gyro resonance for heavier ions are more readily met. Additionally, the pronounced radial component of the crustal magnetic field facilitates vertical ion transport, as detailed in Matta, et al. [10]. Consequently, a vertically expanded distribution of ions in the Martian ionosphere, particularly above the southern highlands, is anticipated. This phenomenon is clearly demonstrated in the statistical dayside distribution of $O^+$ ion density across different altitudes in the geographical coordinates (Fig. 1**B** and Fig. S1).

Contrary to conventional wisdom, the crustal magnetic field aids in atmospheric loss instead of protecting the atmosphere[11,12]. Fig. 1**C** demonstrates a local scenario of ion dynamics above the magnetic anomaly regions. Analogous to the Jeans escape, only limited light ions (e.g., $O^+$) with an outward velocity that is comparable to the escape velocity (5 km/s) can escape along the open magnetic field lines in the absence of ULF waves[10]. With the existence of ULF waves, cold ions get energized through gyro resonance and can overcome the gravitational bounds. Heavier ion species, for instance, meet these conditions at lower altitudes where both magnetic field strength and ion density are relatively higher. In consequence, the impact of ULF waves on the denser, lower ionosphere primarily facilitates the escape of molecular ions, predominantly $O_2^+$ and $CO_2^+$, as proposed in Sakakura, et al. [13].

Yet, to validate the energization and the underlying mechanisms, in-situ observations and analysis are imperative. Recent studies on space plasma physics uncover the wave-ion interaction from a kinetic perspective, marking a critical breakthrough in diagnosing wave-induced ion energization[14-16]. The SupraThermal And Thermal Ion Composition (STATIC) instrument onboard NASA's Mars Atmosphere and Volatile EvolutioN (MAVEN) spacecraft is equipped with a high angular resolution (22.5° × 6°), allowing for precise resolution of ions' pitch angle and gyro phase[17]. Owing to the weak magnetic field and the large mass-charge ratio of ionospheric heavy ions, obtaining observations on a gyro timescale (~21 s for $O^+$ in 50 nT magnetic field) is also within the capability of STATIC's 4-s cadence. In this study, we elucidate the energization mechanism presumably responsible for the catastrophic atmospheric evolution on Mars. Our investigation introduces a case of solar wind-induced ULF waves across the Martian dayside ionosphere, magnetosheath, and solar wind. These ULF waves correlate with a heated energy spectrum (~eV), showcasing the periodic energy dispersion of $O^+$. The presence of ULF modulation and gyro phase bunching stripes confirms the feasibility of the gyro scale interaction between waves and heavy ions. The ongoing interactions are accordingly reproduced by a backward Liouville test particle simulation, which reveals the intricate energization mechanism. This newly identified mechanism offers a plausible explanation for the climate changes during the Hesperian era ~3.7 Ga ago[18]. Our observations are indicative of the prevalent activities on Mars during the early phases of our solar system. Furthermore, severe ion resonant energization and escape are likely still active on habitable exoplanets, particularly those subjected to intensified stellar wind energy and radiation.

**Observation**

Fig. 2 presents an event of solar wind-induced ultra-low frequency (ULF) waves. Variations at the period of ~20 s are evident in ion and magnetic field measurements from the ionosphere to the foreshock (Figs. 2**A-D**). The spacecraft potential measured by the Langmuir Probe and Wave (LPW) is plotted as the dashed grey lines in Figs. 2**A, B**. Planetary ions remain dense and hot until the spacecraft passes through the magnetic pile-up boundary at an altitude of ~850 km (Figs. 2**C, D,** and **E**). The total wave power spectrum in Fig. 2**E** exhibits enhancements around the same frequency (~0.05 Hz) in the ionosphere and solar wind, which reveals the close connection between the ionospheric ULF waves and the foreshock ones. The spacecraft moves equatorward with an outbound trajectory, as shown in Fig. 2**F**. The trajectory in the geographical coordinates (Fig. S2) suggests the spacecraft is traveling through a crustal remnant magnetic field region, consistent with the strong magnetic field observed around UT 21:05:00. Notably, the energy spectrum of the atomic oxygen ion ($O^+$) is revealed to be significantly modulated by the ULF waves where the local $O^+$ gyro frequency coincides with the wave frequency. This coincidence provides us a valuable window to diagnose the ongoing wave-ion resonant interaction in the Martian upper atmosphere.

The evolving wave-ion interaction is further inspected from a microscope view. Fig. 3 displays the analysis of ionospheric $O^+$ measurements accompanied by solar wind-induced ULF waves. The energy spectrum in Fig. 2**B (**and Fig. 3**C)** during this interval can be classified into two parts: the spectrum before 21:07:00 UT is dense but hardly modulated by the weak ULF waves; After this time, the spectrum shows ULF wave-related periodic dispersion signals. The $O^+$ velocity distribution functions within these two segments are given in Figs. 3**A** and 3**B**. Spacecraft potential and plasma bulk velocity are eliminated at each timestamp to resolve the characteristic ion temperature (See Supplementary Material Method: Spacecraft Potential and Ram Velocity Correction on Ion Spectrum, and Fig. S3). Both distributions reveal non-negligible phase space density above the $O^+$ escape energy (~2 eV, corresponding to a speed of 5.0 km/s). Concurrently, the second distribution shows enhanced perpendicular temperature and anisotropy compared to the former. Loss cone signatures are exhibited in both angular-energy spectra. The second spectrum displays asymmetry between the field-aligned and anti-field-aligned sides, hinting at an open-closed field line topology and the escape of the $O^+$.

The following two panels, **D** and **E**, display the angular distribution of the 5.0 – 14.8 eV $O^+$ with respect to pitch angles (PAs) and gyro phases (GPs) (as defined in Method: Ion Gyro Phase and Wave Phase). Periodic variations can be clearly identified in both the PA distributions the GP spectrogram (the latter feature is usually termed as phase bunching in the literature). Considering the instrumental energy sweep rate of 0.25 Hz, the timestamp in panels **D** and **E** is delayed by +750 microseconds. To guarantee reproducibility and minimize potential methodological biases, we generate these spectra based on the raw measurement from the STATIC.

Correspondingly, the wave magnetic field is presented in the last panel. The wave magnetic field is extracted from the measurements through a high-pass filter (with a 0.03 Hz cutoff frequency) and the maximum variant analysis (MVA) method, sequentially. A comprehensive decomposition results of the observed magnetic field time series can be found in Fig. S4. The ratio of the maximum and intermediate eigenvalues is near 5, indicating the observed waves have a linear polarization.

The maximum variation direction is defined as the wave magnetic field polarization direction ($\hat{\mathbf{B}}_{wave}$). The observed ULF waves are highly compressible ($\hat{\mathbf{B}}_{wave} \cdot \hat{\mathbf{B}}_0 \approx 0.96$, where $\hat{\mathbf{B}}_0$ is the direction of the background magnetic field) and were interpreted as fast magnetosonic waves (See Fig. S5 for the ion density response to the ULF waves) in the previous study [6]. Thus, it is reasonable to determine $\hat{\mathbf{E}}_{wave}$ and $\hat{\mathbf{k}}$ as $\frac{\hat{\mathbf{B}}_0 \times \hat{\mathbf{B}}_{wave}}{|\hat{\mathbf{B}}_0 \times \hat{\mathbf{B}}_{wave}|}$ (=[-0.14, 0.99, 0.00]) and $\hat{\mathbf{E}}_{wave} \times \hat{\mathbf{B}}_{wave}$ (=[-0.87, -0.12, 0.48]), respectively (As visualized in Fig. S6)[19]. Here, it is assumed that the wave propagates from the magnetosheath/solar wind to the ionosphere, so we determine $\hat{\mathbf{k}}$ to have a negative radial component.

The wave phase (as defined in Method: Ion Gyro Phase and Wave Phase) is plotted as the solid grey line in Fig. 3**E**. During UT 21:07:00 -21:08:00 interval, the wave phase is locked with the ion gyro phase stripes, indicating the occurrence of the $O^+$ gyro resonance[14-16].

The above analysis procedure is also applicable to the $O_2^+$, as illustrated in Fig. S8. The escape of $O_2^+$ is validated from its asymmetric velocity distribution function. Furthermore, an incomplete gyro phase stripes is observable in this result, indicating a non-resonant response to the waves of the $O_2^+$[20]. However, to maintain clarity, only $O^+$ would be addressed in the main text of this article.

Auxiliary information about the ambient plasma environment is required to understand the undergoing interaction. We include more observations from the Solar Wind Electron Analyzer (SWEA) and the magnetic field model from[21,22]. The modeled magnetic field consisting of the Langlais model and a constant draping term ([-17, -10, -27] nT in MSO coordinates) is added in Fig. 4**A** as a comparison to the observation. The omnidirectional electron energy spectrum is presented in the subsequent panel. A transition from the photoelectron spectrum to the solar wind spectrum arises at UT 21:07:00. This feature suggests that the spacecraft traversed the photoelectron boundary[23].

Utilizing the field line tracing technique based on the magnetic field model, we deduced the topology of the magnetic field. The transition from closed to open field lines at UT 21:07:00, as illustrated in Fig. 4**F**, aligns with the presumed location of the photoelectron boundary inferred from electron measurements. For supplementary clarity, two snapshots of the electron energy spectrum—both preceding and succeeding the photoelectron boundary crossing—within the maximum/minimum pitch angle channels are showcased in Figs. 4**G** and 4**H**. Both the photoelectron peak, observed in the range of approximately 22-27 eV, and the noticeable decline in photoelectron intensity (also known as the 'photoelectron knee') from 60 to 70 eV are discernible in the first two energy spectra at UT 21:05:59. At UT 21:07:29, the characteristics of photoelectrons are only evident in the energy spectrum of electrons moving along field-aligned trajectories (with a pitch angle of 10°). In-situ plasma density observations, derived from LPW and STATIC data, are presented in Fig. 4**C**[24]. During this interval, the Alfvén speed (solid green line in Fig. 4**D**) ranges from ~5 km/s to ~20 km/s and the sonic speed (solid blue line in Fig. 4**D**) that derived from the observed ion moment is less comparable. Their root sum of squares (i.e., the fast magnetosonic speed) is plotted as the solid black line. The dashed black line in Fig. 4**E** shows the filtered magnetic field observation.

Considering the complexity of the detected wave properties and the ambient magnetic field, we will explain the observed ion signature and verify the energization capability with a test particle simulation in the next section. Qualitative discussion about the gyro resonance theory are present in the Method section, from which we can testify the applicability of this mechanism.

**Simulation**

With the above observation, it is feasible to construct the wave and ambient magnetic field with a series of empirical analytic functions (See Method: Test Particle Setup). Thereafter, we conduct a backward Liouville test particle simulation to investigate the wave-ion interaction. The backward tracing is stopped as the particle impacts the local magnetic pile-up boundary layer (at an altitude of 850 km) or exits the wave field (at an altitude of 500 km, corresponding to a wave electric field amplitude of ~0.01 mV/m). The initial distribution of $O^+$ at the upper and lower boundary (500 km) is set to 0 and an isotropic Kappa distribution with $\kappa = 6$ and $T_0 = 2.2$ eV, respectively. The comparison between the Kappa model and observation is presented in Fig. S9.

The simulation results are presented in Fig. 5. Fig. 5**A** presents the simulated energy spectrum with the STATIC's energy and time resolution. Values in each bin are averaged from 9216 test particles with varying release times ($\Delta t = 0.5$ s), pitch angles ($\Delta PA = 7.5°$), and gyro phases ($\Delta GP = 7.5°$).

Figs. 5**B-D** is in the same format as Figs. 3**D-F**. Impressively, our simulation faithfully reproduces the observation characteristics, including energy dispersion stripes, pitch angle modulations, and gyro phase bunching stripes in detail. The characteristic ion trajectories in the Energy-$\zeta$ ($\zeta := \phi_B - \phi_{ion}$) and Altitude-$\zeta$ phase space, as shown in Fig. S10, manifest the occurrence of the resonant energization (from the energy of <1 eV to that of > 10 eV) and ion escape (from the altitude of 500 km to that of 800 km). Following the same methodology, it is feasible to model the $O_2^+$ response to the waves, which is presented in Fig. S11.

Via test-particle simulation, we have confirmed the ongoing energy interchange between solar wind-induced ULF waves and heavy ions in the Martian ionosphere. Those energized ions will further undergo the thermalization process or directly escape the ionosphere, subjecting to the field line topology. These two processes are evidenced by the observed hot $O^+$ energy spectrum with a single side loss cone, as shown in Figs. 3**A-B**.

## Discussion

This research suggests that the solar wind's energy can be transferred to planetary heavy ions via electromagnetic waves. Electromagnetic waves with typical parameters in our case (an amplitude of ~ 5 nT and a phase velocity of ~10 km/s) carry with them a Poynting flux of $\frac{1}{2}\left(V_P \frac{B_w^2}{\mu_0}\right) \approx$ 6.2×10$^{11}$ eV/s/m$^2$. Therefore, by assuming a cross-section of $\pi R_M^2$ (1/4 of the planetary surface area, where $R_M$ is the Mars Radius, 3393 km), the total energy transmit rate can be estimated to be ~ 2.2 × 10$^{25}$ eV/s. A characteristic escape rate for oxygen atoms can hence be derived to be ~1.1 × 10$^{25}$ s$^{-1}$ by dividing the energy input rate by $O^+$ escape energy. This quantity is close to the previous statistical result on the mean oxygen ion escape rate (~1.7 × 10$^{25}$ s$^{-1}$)[13,25]. However, our estimation may just represent an upper limit since we assume all wave energy is converted to the ion kinetic energy; the gained energy may further be converted to thermal energy through other thermalization processes. The heating process is validated by the ~eV temperature energy spectrum with a high energy tail during UT 21:04:00 – 21:08:30. While the ion energy spectrum before UT 21:03:30 is significantly cooler, revealing a less-heated ionosphere population (Fig. 2**B**).

It is essential to note that the above wave parameters are derived from ionospheric observations rather than from the magnetosheath or solar wind. Caution is advised when interpreting the energy conversion from the solar wind to these waves, which is not the primary focus of this investigation. Fig. S12 showcases the power spectra of proton pressure, proton density, and the magnetic field around 0.05 Hz. Their common ~20 s periodicity suggests that neither can be ruled out as contributors to transmit or excite ULF waves in the ionosphere.

Additionally, this scenario provides insights into the preferential conditions that might lead to ionospheric escape. The planet's crustal magnetic field also plays a pivotal role: by regulating the gyro frequency, it introduces geographic variations in the rate of loss, as partially evidenced by Fig. 1**B** and Fig. S1. Outside the magnetic anomaly region, the magnetic field intensity is nearly half (Fig. S1). A Doppler velocity near the escape velocity ($V_\parallel \geq V_P[1 - \Omega/\omega] \approx 5$ km/s) is at minimum required to compensate for the mismatches in the resonant condition. This preference revolutionizes our understanding of the influence of planetary magnetic fields on solar-planet interactions. The resonance conditions further imply a species preference: ignoring the negligible Doppler effect caused by the thermal motion, ions with a higher mass-charge ratio are more likely to resonate at lower altitudes where the crustal magnetic field is stronger. This might explain the $CO_2^+$-rich plume escape events documented in Sakakura, et al. [13]. These insights, still within the scope of MAVEN, certainly deserve further exploration. The hemisphere closer to the quasi-parallel bow shock is more directly influenced, thereby harnessing a larger share of upstream energy.

The substantiation of wave-participated energy transfer path provides a novel perspective to evaluating climate evolution. This perspective diverges from traditional views by presenting a theory that directly connects solar wind energy with ionospheric loss. Such a linkage supports the rapid water and atmospheric loss during the Hesperian era. A total mass loss over 3.9 Ga can be estimated up to hundreds of mbar, assuming the loss rate is linearly dependent on the solar wind ram pressure (See Method).

Beyond Mars, exoplanets within the prevalent M-type dwarf stellar systems might also exhibit a stellar wind–wave–ionosphere interaction. These stars, characterized by their relatively lower radiation temperatures and more active atmospheres, foster interstellar environments with a heightened energy density from the stellar wind, especially around the habitable zones.

**Methods**

Ion Gyro Phase and Wave Phase

The wave phase of the observed linear wave can be derived as $\phi_B := \arcsin\left(\frac{B_{wave}}{B_{amp,wave}}\right)$, where $B_{wave}$ is the filtered wave magnetic field along $\hat{\mathbf{B}}_{wave}$ direction and $B_{amp,\,wave}$ is the wave amplitude obtained from the Hilbert transform of $B_{wave}$. Such a linear polarized wave can be regarded as a superposition of a pair of left-handed and right-handed polarized waves with the same amplitude and wave vector. Our definition of $\phi_B$ is equivalent to the polar angle of the left-handed component of the $\mathbf{B}_{wave}$ in the $\hat{\mathbf{B}}_{wave}$-$\hat{\mathbf{E}}_{wave}$ plane.

Correspondingly, the ion gyro phase ($\phi_{ion}$) is defined as the polar angle of the ion perpendicular velocity in the $\hat{\mathbf{B}}_{wave}$-$\hat{\mathbf{E}}_{wave}$ plane. The decomposition of the linear polarized waves and the definitions of $\phi_B$ and $\phi_{ion}$ are visualized in Fig. S7.

Test Particle Setup

To model the wave-ion interaction, we employ numerical approximation to reconstruct the observed wave properties as the test particle simulation setup. As shown in Figs. **4D** and **4E**, the observed wave phase speed (fast magnetosonic speed) and the wave amplitude increase as the spacecraft move from the altitude of ~500 km to that of ~800 km. Hence, both the wave phase speed and amplitude are assumed to be exponentially dependent on the altitude in our numerical reconstruction:

$$V_P = V_{p0} \exp\left(\frac{r - r_0}{H_1}\right)$$

$$\mathbf{E}_{wave} = \hat{\mathbf{E}}_{wave} \cdot E_0 \exp\left(\frac{r - r_0}{H_2}\right) \cos(\int \mathbf{k} \cdot d\mathbf{r} - \omega t)$$

Where $V_P$, $\mathbf{E}_{wave}$ is the wave phase velocity and the wave electric field, respectively. $V_{p0}$ (20 km/s) and $E_0$ (0.15 mV/m) are the reference values at a fixed altitude of 800 km. And, the e-folding height for the phase velocity and wave electric field amplitude is denoted as $H_1$ (240 km) and $H_2$ (120 km). The wave angular frequency $\omega$ is set as $2\pi/21.5$ rad/s in the MSO reference frame. While the wave frequency (0.052 Hz, as shown in Fig. 4**I**) observed by the spacecraft is slightly higher due to the Doppler effect. The wave magnetic field B$_{wave}$ can be self-consistently derived from the Faraday's law ($\mathbf{B_{wave}} = \frac{\nabla \times \mathbf{E_{wave}}}{i\omega}$). All the aforementioned parameters are estimated from the observation shown in Fig. 4. The modeled phase speed and wave magnetic field are shown as the dashed red lines in Fig. 4**D** and 4**E**. The reconstructions can approximately approach the observation and are suitable for analyzing of the wave-ion interaction within the region of interest.

We implement it with the 4$^{th}$-order Runge-Kutta method, with a time step of 10 microseconds.

More details about the gyro resonance theory

In the studied event, the ionospheric ULF waves have a large perpendicular wave vector ($k_\perp$) and the phase bunching condition requires further validation. The cosine dependence on the ion gyro phase of the wave phase can be eliminated after gyro averaging [26]. And, the trapping frequency ($\omega_{tr}$) has to be modified as $\sqrt{J_0(k_\perp \rho)}\omega_{tr0}$, where $\omega_{tr0} = \sqrt{k_\parallel v_\perp(qB_{wave}/m)}$ is the classical trapping frequency, $J_0$ is the 0-th order Bessel function, and $\rho := mv_\perp/qB_0$ is the ion gyro radius. Thus, the phase bunching is still possible as long as $J_0(k_\perp \rho) > 0$. While, the ion magnetic moment and energy will oscillate at the wave frequency, corresponding to the energy dispersion stripes we observed. Similarly, the presence of the right-hand polarized wave component also leads to an ion energy modulation above the wave frequency without violating the phase bunching. In addition, the up-flowing ions sense a growing resonance island due to the vertical gradient of the wave amplitude and its phase space trajectory should be elongated along the magnetic moment axis or the energy axis. Overall, gyro resonance is fully applicable for interpreting the ion energization phenomenon in the Martian magnetosphere theoretically.

Kappa Distribution in the gravitational field

The model phase space density (f) is given as follows:

$$f(r, v) = \frac{n(r)}{2\pi(\kappa w_\kappa^2(r))^{3/2}} \frac{\Gamma(\kappa+1)}{\Gamma(\kappa-1/2)\Gamma(3/2)} \left(1 + \frac{v^2}{\kappa w_\kappa^2(r)}\right)^{-\kappa-1}$$

Where r is the altitude, $n(r) = n(r_0)\left[1 + \frac{R(r)-R(r_0)}{\kappa w_\kappa^2(r_0)}\right]^{-\kappa+1/2}$ is the number density of particle, $w_\kappa(r) = w_\kappa(r_0)\left[1 + \frac{R(r)-R(r_0)}{\kappa w_\kappa^2(r_0)}\right]^{1/2}$ is the thermal velocity of particle, $w_\kappa(r_0) = \left(\frac{2\kappa-3}{\kappa}\frac{k_B T_0}{m}\right)^{1/2}$ is the initial value of thermal velocity converted from Maxwellian temperature $T_0$, $R(r) = -\frac{GMm}{r}$ is particle's gravity potential [27].

Spacecraft Potential and Ram Velocity Correction on Ion Spectrum

1) Modify the ion energy by the observed spacecraft potential: $E' = E + q\varphi$

2) Compute the ion speed according to the modified energy: $v' = \sqrt{\frac{2E'}{m}}$

3) Multiply the ion speed with instrumental view direction (**n**): $\mathbf{v}' = \mathbf{n}v'$

4) Subtract the ram velocity of the spacecraft: $\mathbf{v}'' = \mathbf{v}' - \mathbf{v}_{ram}$

The correction procedures are visualized in Fig. S3.

Estimation of Total Mass Loss within 3.9 billion Years

Assuming a direct proportionality between the escape rate and the solar wind dynamic pressure, we can express the escape rate (Q) at any given time (T) as follows:

$$Q(T) = Q_0 \cdot (T/T_0)^\alpha$$

Where $Q(T)$ is the escape rate at time T, $T_0$ represents the present age of the Sun (4.6 billion years), $Q_0$ is the contemporary oxygen escape rate, set at $1.7 \times 10^{25}$ s$^{-1}$ [25], $\alpha$ denotes the power-law index (set at -2.64) correlating the solar wind dynamic pressure with time (solar wind dynamic pressure $\propto T^\alpha$) [28,29].

To calculate the total mass loss ($\Delta M$), we integrate the escape rate over the specified time range (3.9 Ga to present):

$$\Delta M = M_O \cdot \int_{0.7}^{4.6} Q(T)dT = M_O Q_0 T_0^{-\alpha} \cdot \int_{0.7}^{4.6} T^\alpha dT = \frac{M_O Q_0 T_0^{-\alpha}}{(\alpha + 1)} \cdot T^{\alpha+1}\Big|_{0.7}^{4.6} = 8.4 \times 10^{17} \text{kg}$$

Where $M_O$ is the mass of an atomic oxygen.

**Acknowledgments:**

This work was supported by the National Natural Science Foundation of China 42230202 (QGZ) and the Major Project of Chinese National Programs for Fundamental Research and Development 2021YFA0718600 (QGZ).

**Funding:**

National Natural Science Foundation of China 42230202 (QGZ)

Major Project of Chinese National Programs for Fundamental Research and Development 2021YFA0718600 (QGZ)

**Author contributions:**

Conceptualization: QGZ, XZZ, JTZ

Methodology: JTZ, JHL

Investigation: JTZ, ZYL, SW, XZZ

Visualization: JTZ

Funding acquisition: QGZ

Project administration: QGZ

Supervision: QGZ, XZZ, SW, WHI

Writing – original draft: JTZ

Writing – review & editing: JTZ, QGZ, ZYL, SW, XZZ, WHI, CY, YXH, RR, AD, SYF, HZ, YFW

**Competing interests:** Authors declare that they have no competing interests.

**Data and materials availability:** MAVEN STATIC, MAG, SWEA, and LPW data are available from the MAVEN Science Data Center (https://lasp.colorado.edu/maven/sdc/public/). The spherical harmonic coefficient of the magnetic field model used in this study (Langlais 2019) can be found in the supporting information of [22]. Calibrated ion density data is derived by [24] and provided in [30] (SPEDAS). The test particle simulation code is available from the corresponding author upon request.

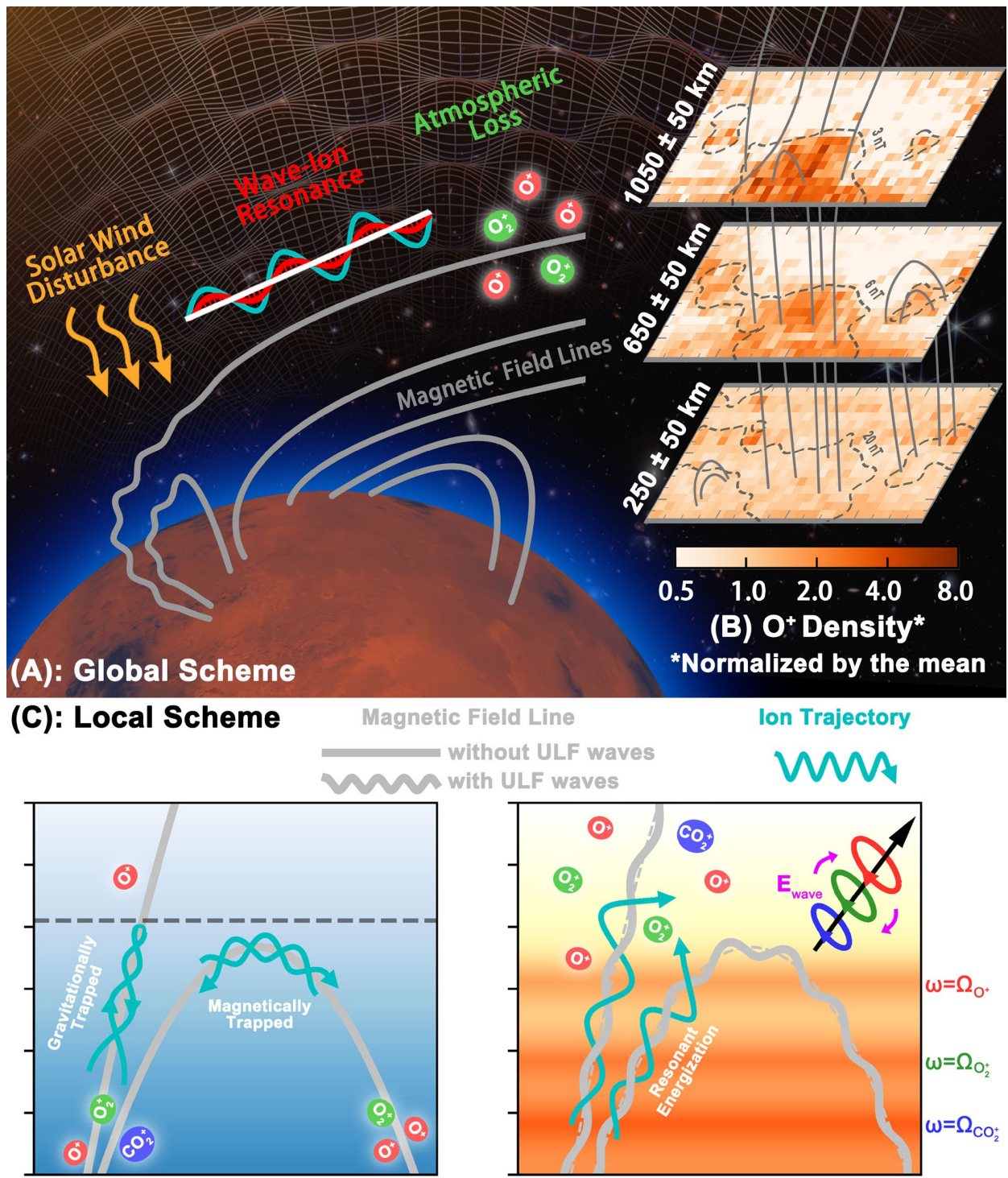

**Fig. 1. Schematic Illustration of Solar Wind-Induced Atmospheric Loss via ULF Wave-Ion Gyro Resonance.** (**A**) Global Energy Transfer Scenario: Periodic structures formed in the Martian bow shock's upstream region (solar wind energy) give rise to ULF perturbations within the ionosphere (ULF wave energy), energizing the cold trapped ions through gyro resonance (particle energy). (**B**) Spatial Distribution of $O^+$ Density: Shown at altitudes of $250 \pm 50$ km, $650 \pm 50$ km, and $1050 \pm 50$ km (from the bottom to the top), with oxygen ion density normalized to their respective mean values within each altitude band. Smoothed modeled magnetic field

intensity is illustrated by the dashed grey contour lines (20 nT, 6 nT, 3 nT, respectively). (**C**) Local Particle Escape Dynamics: In the absence of ULF waves (left panel), only a few ionized particles exceeding escape velocity can leave the ionosphere. With ULF waves (right panel), charged particles are sufficiently energized by gyro resonance within the wave field, facilitating their ionospheric escape. Resonant interactions for $O^+$, $O_2^+$, and $CO_2^+$ ions occur at varying altitudes, reflecting their mass differences.

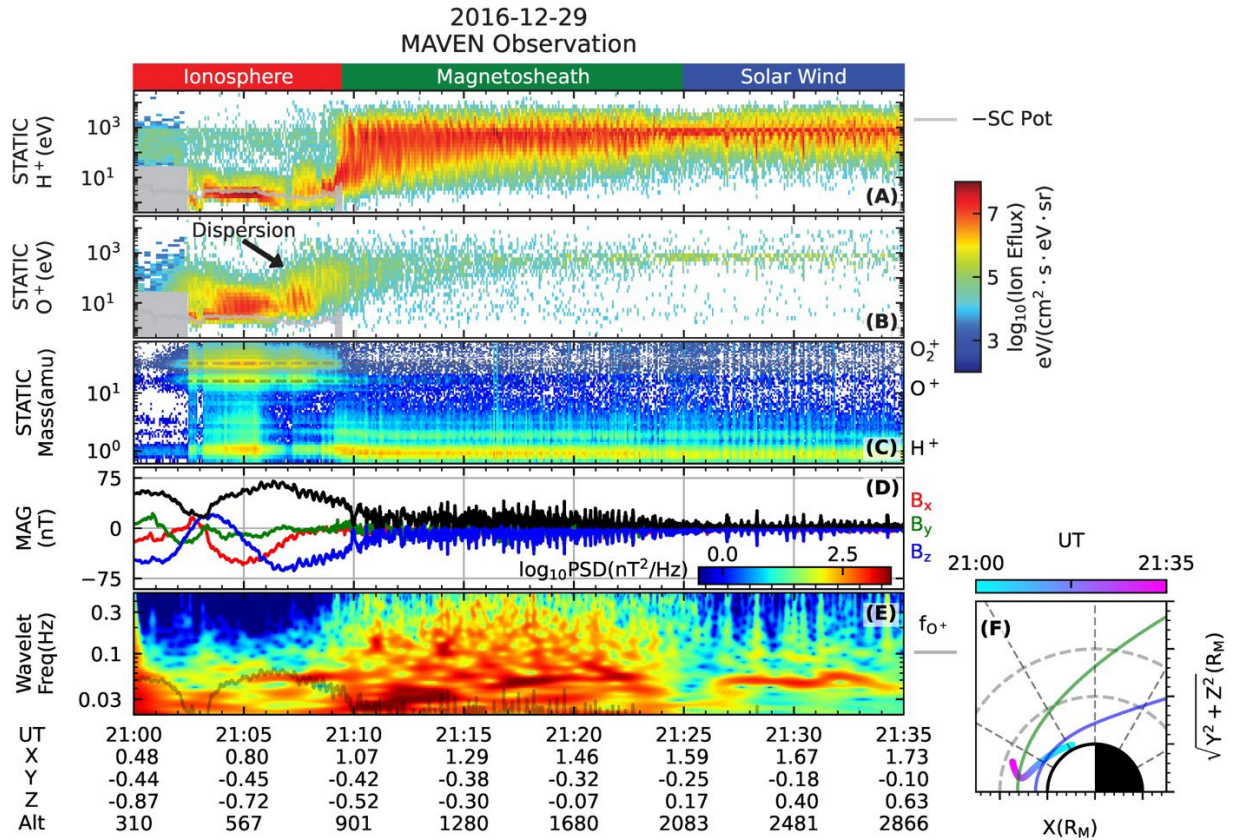

**Fig. 2. Overview of MAVEN's Ion and Magnetic Field Observations on December 29, 2016.** (**A**) Proton (H$^+$) Energy Spectrum from STATIC Instrument: The shaded area in the lower-left indicates the minimum detectable energy, with the solid grey line depicting the spacecraft's negative potential (SC Pot) as determined by Langmuir Probe and Wave data. (**B**) STATIC's Atomic Oxygen Ion (O$^+$) Energy Spectrum. (**C**) STATIC Mass Spectrum: The masses of O$^+$ and O$_2^+$ ions are distinctly identified by two horizontal dashed lines. (**D**) Magnetometer Data: Time series of the three components (B$_x$ in red, B$_y$ in green, B$_z$ in blue) in the Mars-Sun-Orbital (MSO) coordinate system. (**E**) Total Wave Power Spectrum: Derived using Morlet wavelet analysis, with the O$^+$ gyro frequency represented by a dashed grey line. (**F**) Spacecraft Trajectory: Shown in the X-$\sqrt{Y^2 + Z^2}$ plane of the MSO system, including modeled magnetic pile-up boundary (in blue) and bow shock (in green) as referenced from [31]. [The wavelet power density shown in panel E should be multiplied by a constant normalization factor. we find it after the submission, and we will revise it during the revision.]

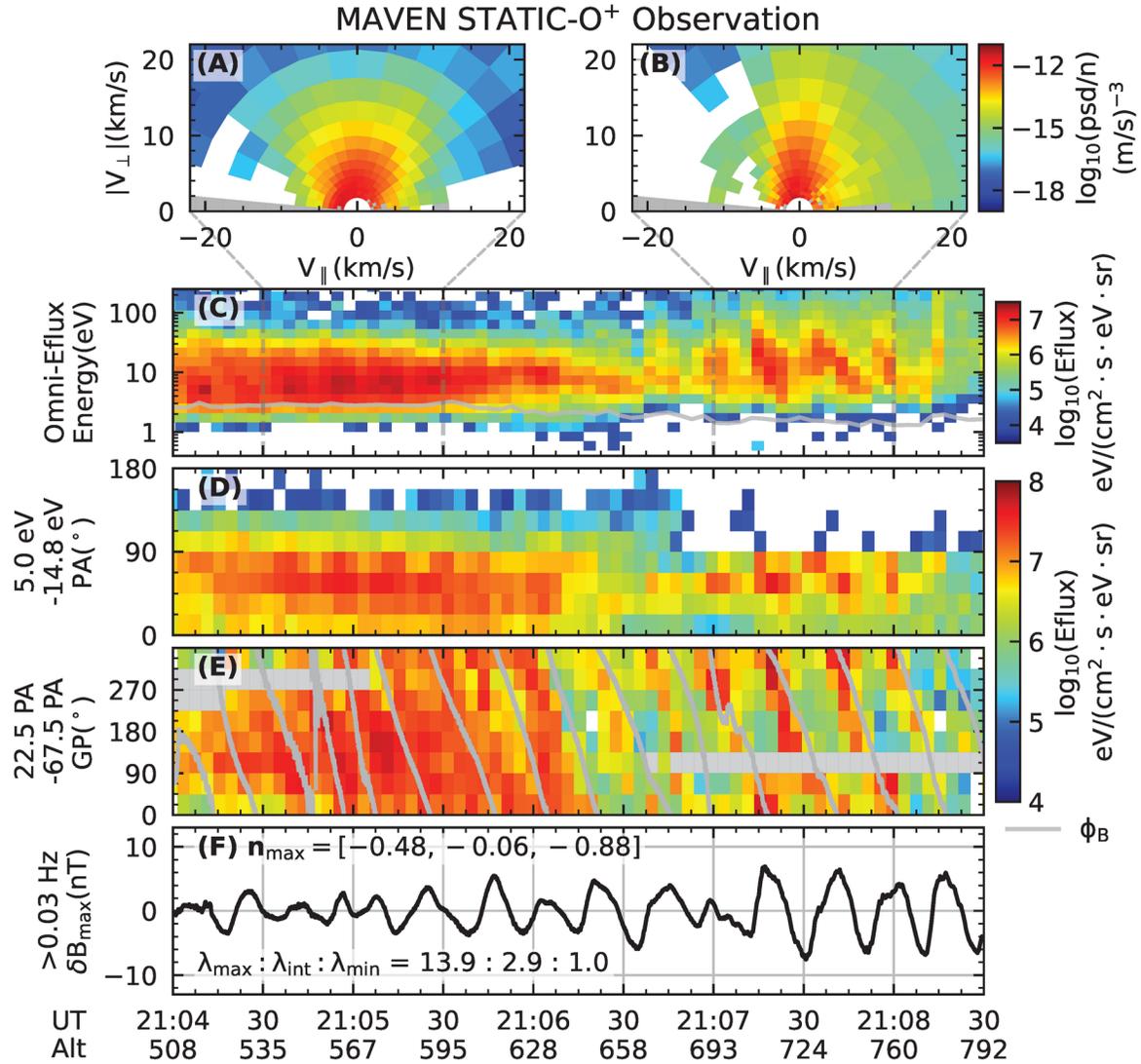

**Fig. 3. Kinetic features of $O^+$.** (**A**) Two-dimensional velocity distribution function (i.e., phase space density divided by the number density) of $O^+$ between UT 21:04:30 and UT 21:05:30 with the correction for spacecraft potential and $O^+$ bulk velocity. (**B**) Same as panel A, but between UT 21:07:00 and UT 21:08:00. (**C**) Omni-directional energy spectrum (in energy flux) from the raw STATIC-c6 data. (**D**) Pitch angle spectrum of 5.0 eV - 14.8 eV $O^+$. (**E**) Gyro phase spectrum of 5.0 eV – 14.8 eV, 22.5°-67.5° pitch angle $O^+$. Wave phase($\phi_B$) is plotted by the solid grey line. (**F**) Maximum variant component of filtered (high pass filter with a cutoff frequency of 0.03 Hz) magnetic field series.

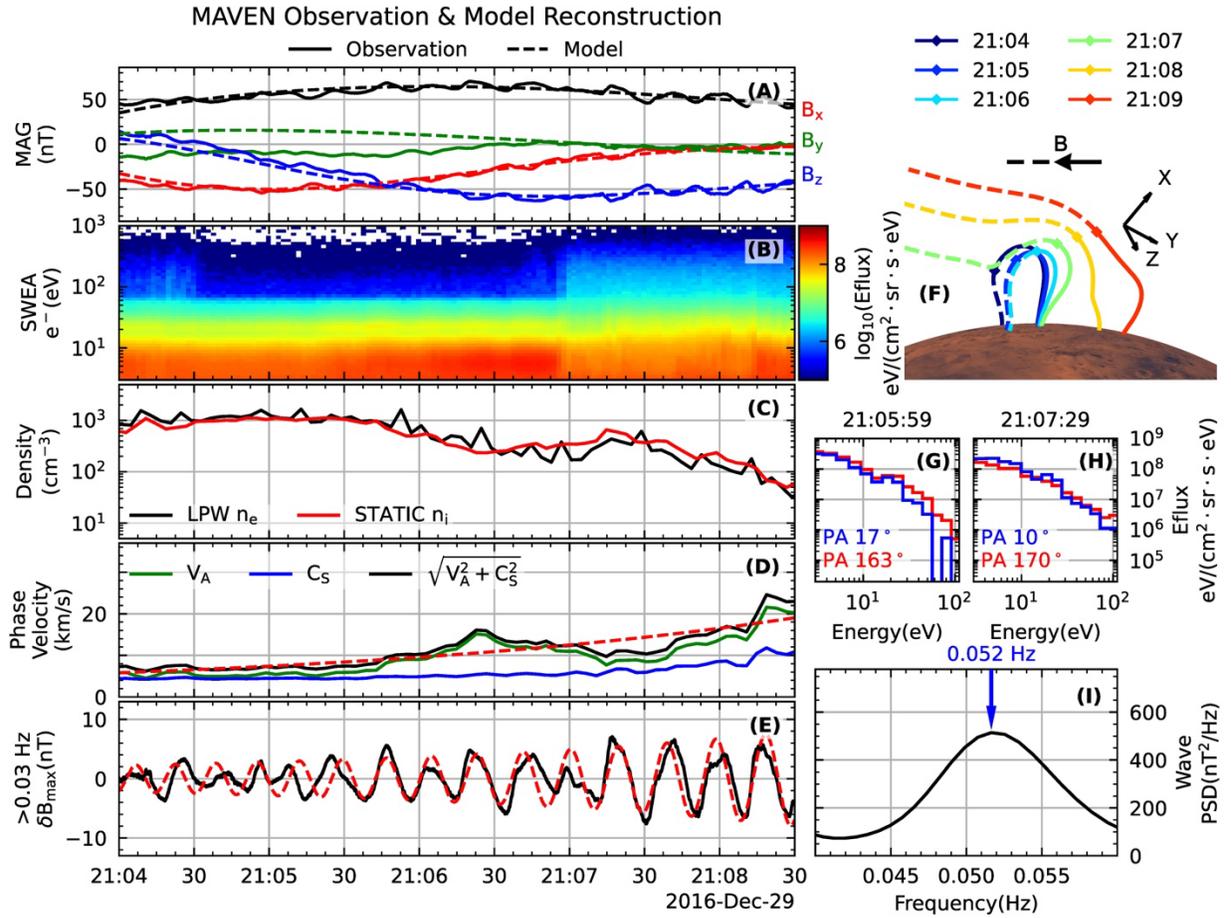

**Fig. 4. MAVEN's magnetic field and electron observation during the same interval as Fig. 3.** (**A**) Magnetic field time series in MSO coordinates. Solid lines: Magnetometer's observation; Dashed lines: Modeled magnetic field (the Langlais 2019 model with an offset of (-17, -10, -27) nT along MSO X, Y, Z direction)[22] (**B**) SWEA electron energy spectrum (energy flux). (**C**) Ion (STATIC, solid red line) and electron (LPW, solid black line) density. (**D**) Observed Alfvén ($V_A$, solid green line) and sonic speed ($C_S$, solid blue line). The fast magnetosonic speed (the root sum of square of $V_A$ and $C_S$) and the modeled wave phase speed is plotted as black and red lines, respectively. (**E**) Same as Fig. 3F. The dashed red line represents the modeled wave signal. (**F**) Topology of magnetic field determined via modeled field line tracing. Basis in MSO coordinates are denoted as black arrows. The magnetic field direction extends from the solid line towards the dashed line. (**G**) Electron energy spectrum within minimum/maximum pitch angle channel at UT 21:05:59 (**H**) and UT 21:07:29. (**I**) Mean wave power spectrum density of the observed magnetic field between UT 21:04:00 and UT 21:08:30. [The wavelet power density shown in panel I should be multiplied by a constant normalization factor. we find it after the submission, and we will revise it during the revision.]

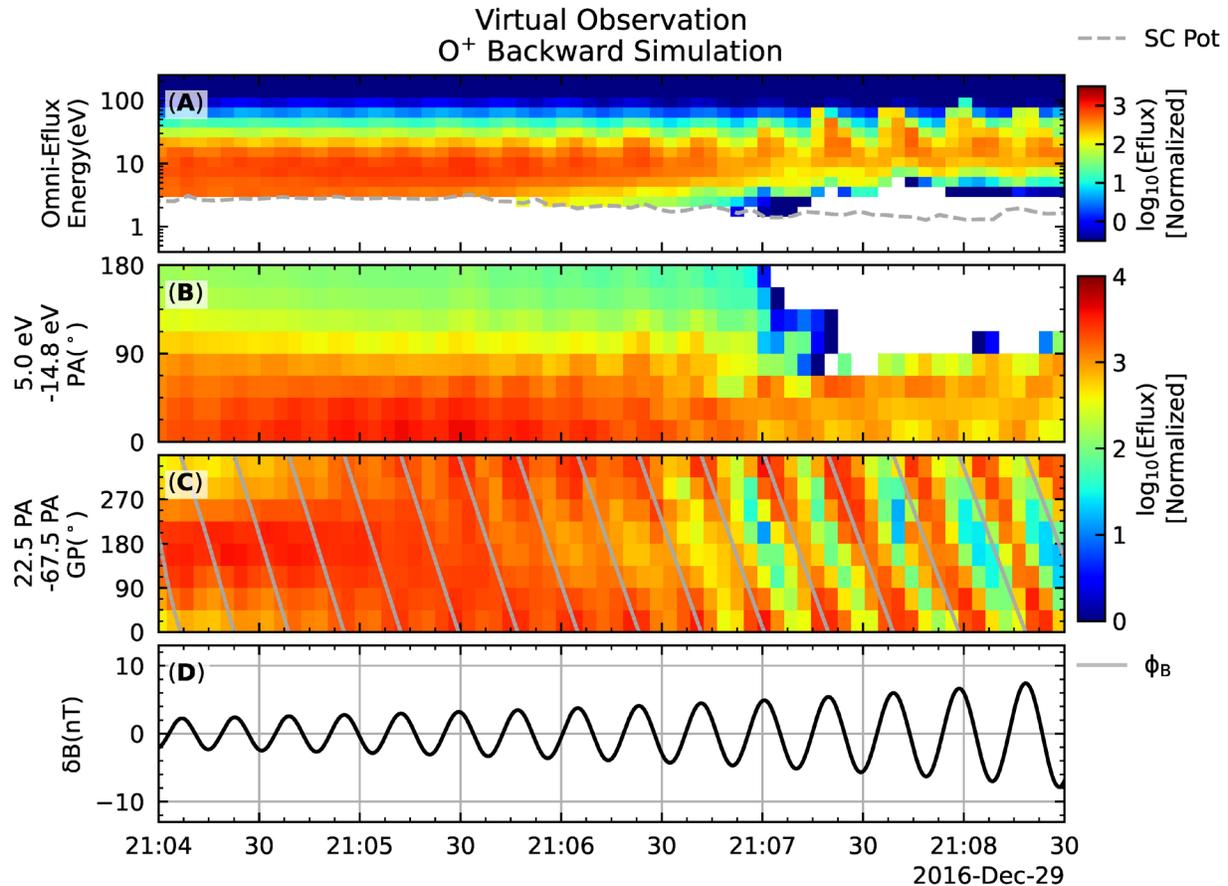

**Fig. 5. Virtual observation of backward Liouville simulation for O$^+$.** (**A**) Omni-directional energy flux in an arbitrary unit. (**B**) Pitch angle spectrum of 5.0 eV -14.8 eV O$^+$. (**C**) Pitch angle spectrum of 5.0 eV - 14.8 eV O$^+$. (**D**) Gyro phase spectrum of 5.0 eV – 14.8 eV, 22.5° -67.5° pitch angle O$^+$. Wave phase ($\phi_B$) is plotted by the solid grey line. (**F**) Modeled wave magnetic field.

# Supplementary Materials for

## On Energization and Loss of the Ionized Heavy Atom and Molecule in Mars' Atmosphere

J. -T. Zhao[1], Q. -G. Zong[2,1*], Z. -Y. Liu[3], X. -Z. Zhou[1], S. Wang[1], W. -H. Ip[4], C. Yue[1], J. -H. Li[1], Y. -X. Hao[5], R. Rankin[6], A. Degeling[7], S. -Y. Fu[1], H. Zou[1], and Y. -F. Wang[1]

[1] Institute of Space Physics and Applied Technology, Peking University, Beijing, China.

[2] Key Laboratory of Lunar and Planetary Sciences, Macau University of Science and Technology (MUST), Macau, China.

[3] Institut de Recherche en Astrophysique et Planétologie, CNES-CNRS-Universite Toulouse III Paul Sabatier, Toulouse, France.

[4] Institute of Astronomy, National Central University, Jhongli, Taiwan.

[5] Max Planck Institute for Solar System Research, Göttingen, Germany.

[6] Department of Physics, University of Alberta, Edmonton, AB, Canada.

[7] Shandong Provincial Key Laboratory of Optical Astronomy and Solar-Terrestrial Environment, Institute of Space Sciences, Shandong University, Weihai, China.

*Corresponding author. Email: qgzong@pku.edu.cn

**The PDF file includes:**

    Figs. S1 to S12
    References

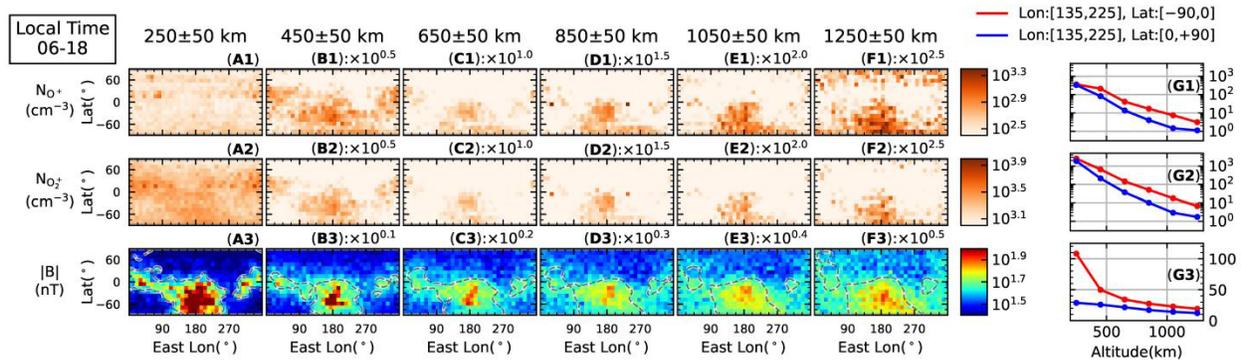

**Fig. S1. Geographical distribution of dayside $O^+$ density, $O_2^+$ density, and magnetic field intensity derived from MAVEN-STATIC d1 data. (A1-F1)** $O^+$ density distribution within the altitude range of 250 ± 50 km, 450 ± 50 km, 650 ± 50 km, 850 ± 50 km, and 1050 ± 50 km, respectively. **(A2-F2)** $O_2^+$ density distribution within the altitude range of 250 ± 50 km, 450 ± 50 km, 650 ± 50 km, 850 ± 50 km, and 1050 ± 50 km, respectively. **(A3-F3)** Distribution of the observed magnetic field intensity at altitudes of 250 ± 50 km, 450 ± 50 km, 650 ± 50 km, 850 ± 50 km, and 1050 ± 50 km is presented. Grey contour lines illustrate the smoothed Langlais 2019 modeled magnetic field intensities of 20, 12, 6, 4, 3, and 2 nT at each corresponding altitude band [1]. **(G1, G2, G3)** Altitude dependence of the $O^+$ density, $O_2^+$ density, and the magnetic field intensity for the magnetic anomaly region (red) and the opposite hemisphere (blue).

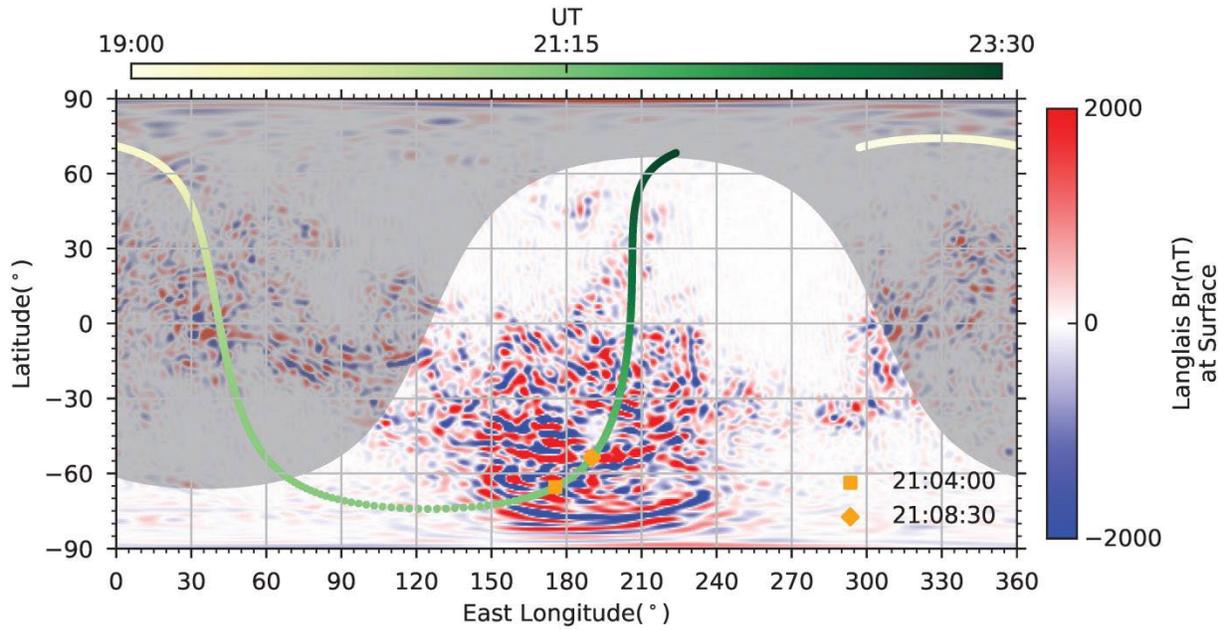

**Fig. S2. MAVEN's trajectory in geographical coordinates system.** The spacecraft's trajectory is projected onto the longitude-latitude plane. The background map is color-graded based on the radial component of the Langlais 2019 model at the planetary surface [1].

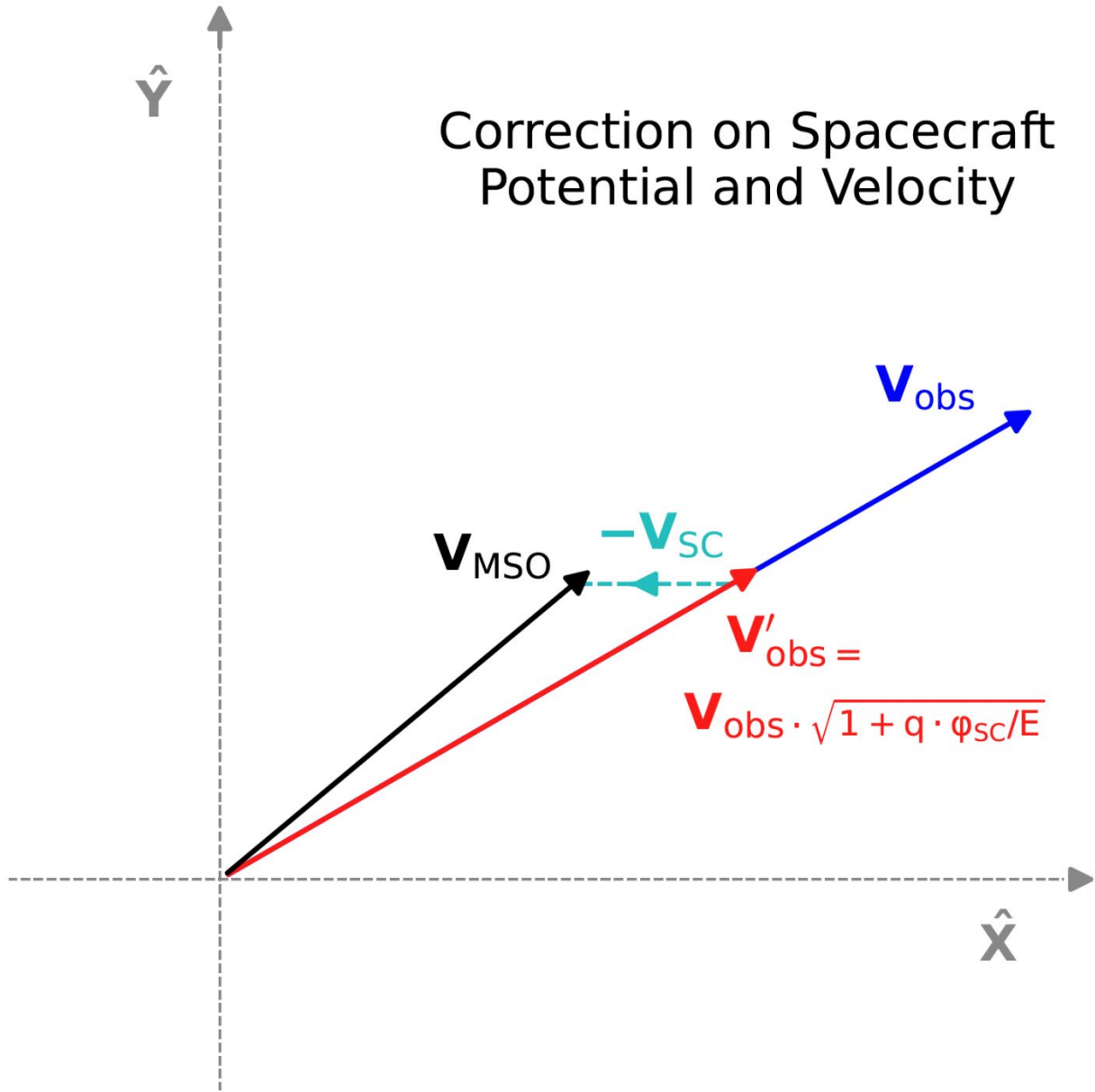

**Fig. S3. Visualization of the spacecraft potential and ram velocity correction.** Blue arrow ($\mathbf{V}_{obs}$) denotes the velocity that directly derived from the STATIC measurement and in the spacecraft reference frame. Red arrow ($\mathbf{V}'_{obs}$) denotes the velocity that corrected by the spacecraft potential in the spacecraft frame. After subtracting the spacecraft ram velocity, the origin velocity of the ion in the MSO frame is derived as $\mathbf{V}_{MSO}$ (black arrow).

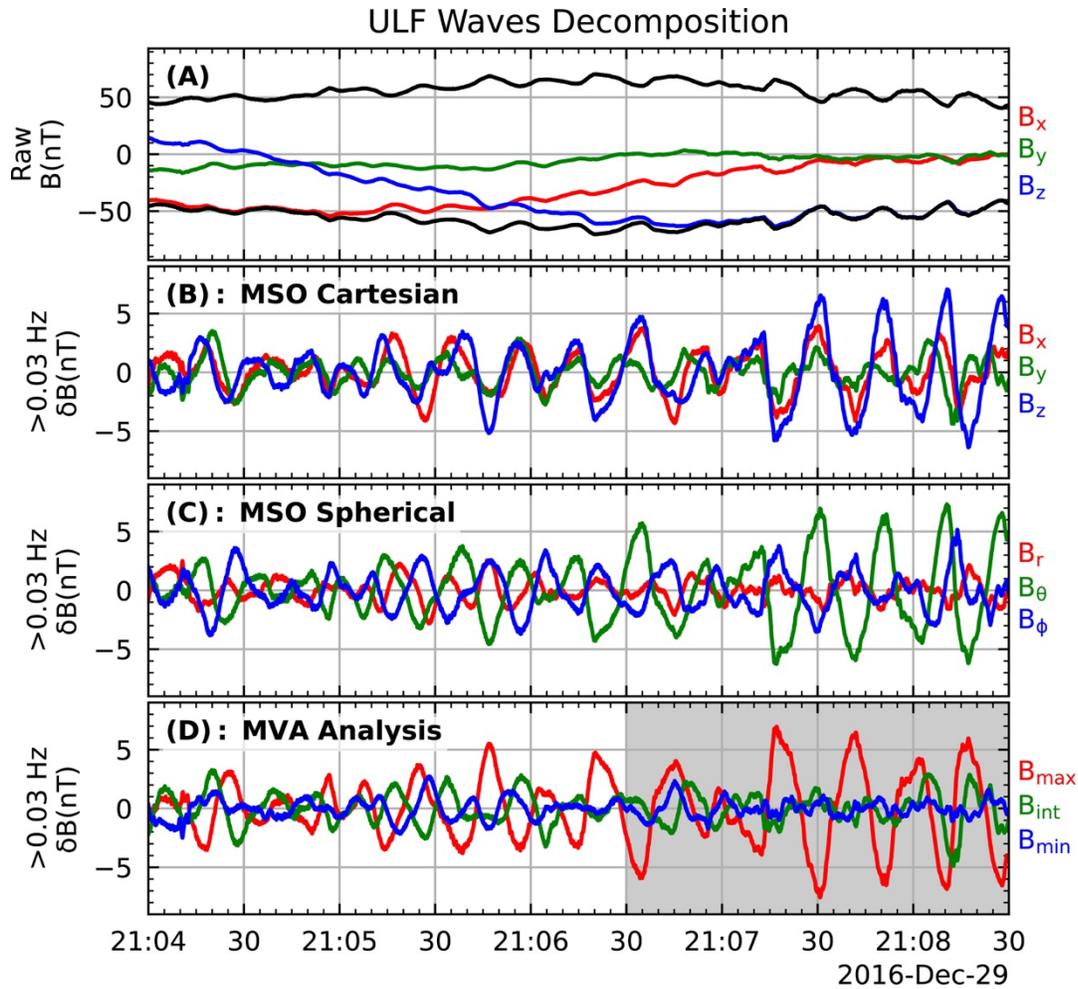

**Fig. S4. Decompositions of observed ULF waves in the Martian ionosphere.** (A) Raw magnetic field data in the MSO coordinates. The wave signals are extracted by a 0.03 Hz high-pass filter from the raw data. (B) Wave magnetic field in cartesian MSO coordinates. (C) Wave magnetic field in spherical MSO coordinates. (D) Wave magnetic field in minimum variance coordinates. The shaded area indicates the time range (UT 21:06:30 – UT 21:08:30) we select for wave orientation analysis (including MVA analysis and the calculation of the mean magnetic field). The maximum and minimum variation direction in the MSO coordinates is $[-0.48, -0.06, -0.88]$ and $[-0.70, 0.65, 0.30]$. It should be noted that the MVA result almost does not change if we select the magnetic field measurement between UT 21:04:00 – UT 21:08:30 in MVA analysis (the corresponding maximum variation direction is $[-0.51, -0.15, -0.85]$).

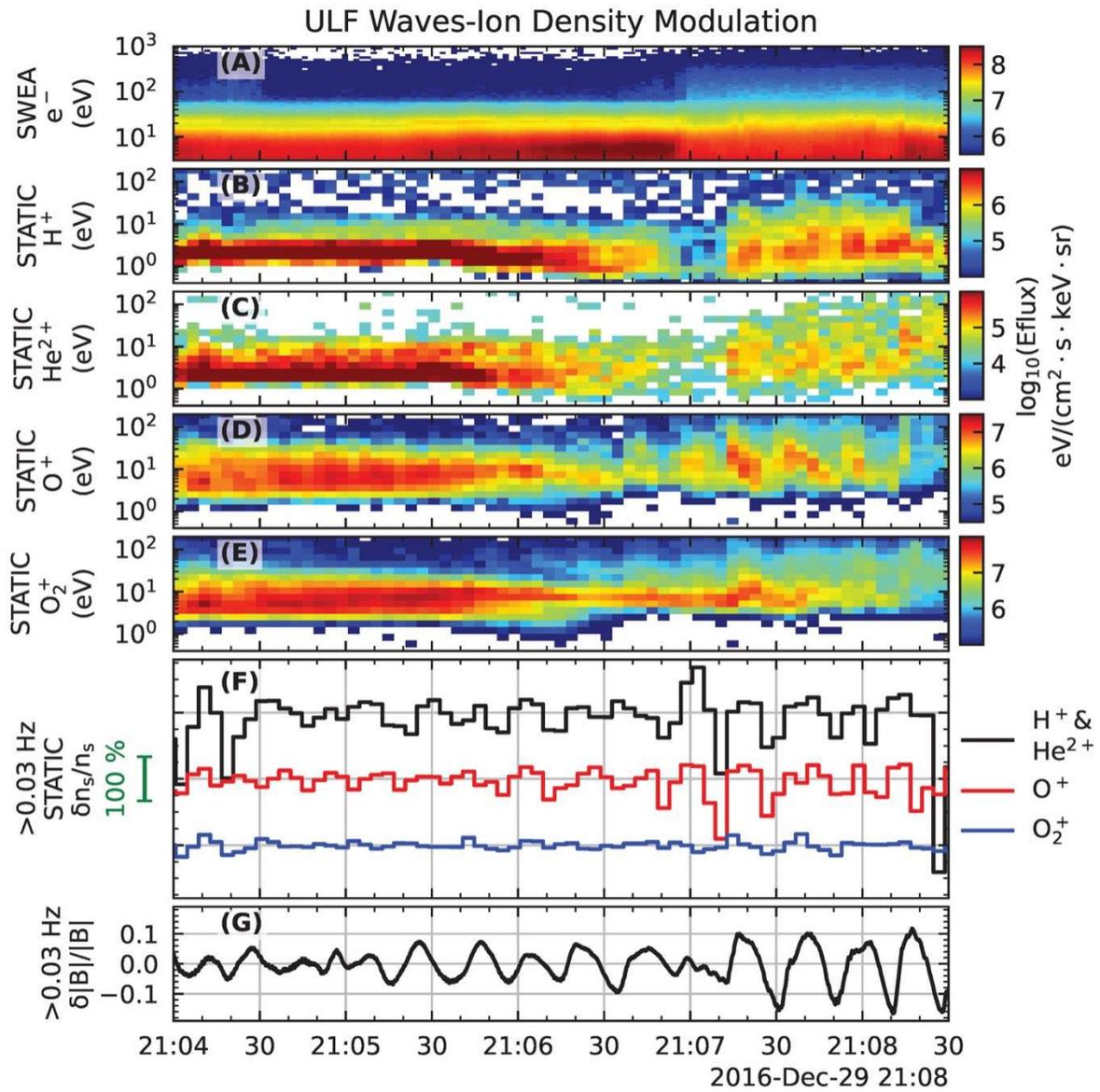

**Fig. S5. Modulation of Ion Density by ULF Waves.** (**A**) SWEA measurements of the electron energy spectrum. (**B-E**) STATIC measurements of the energy spectra for $H^+$, $He_2^+$, $O^+$, and $O_2^+$ ions, respectively. (**F**) Fluctuations in ion density for each species. (**G**) Variations in magnetic field intensity.

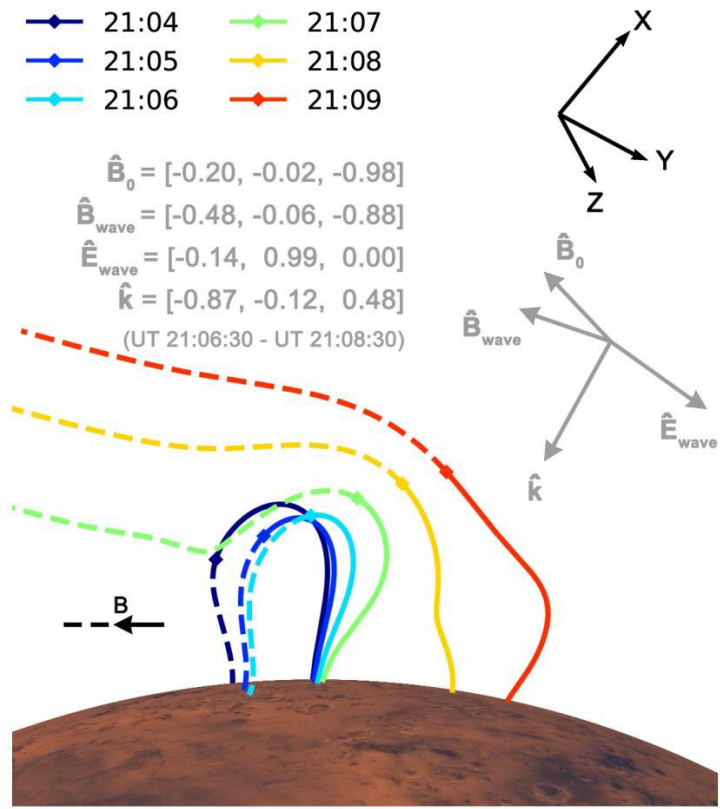

**Fig. S6. Visualization of the wave orientation.** Colored lines present the traced magnetic field lines from the Langlais 2019 model (with a constant draping term). Grey arrows show the directions of the background magnetic field ($\hat{\mathbf{B}}_0$, determined from the mean magnetic field), wave magnetic field ($\hat{\mathbf{B}}_{wave}$, obtained from MVA analysis), wave electric field ($\hat{\mathbf{E}}_{wave}$, derived as $\frac{\hat{\mathbf{B}}_0 \times \hat{\mathbf{B}}_{wave}}{|\hat{\mathbf{B}}_0 \times \hat{\mathbf{B}}_{wave}|}$), and wave vector ($\hat{\mathbf{k}} = \hat{\mathbf{E}}_{wave} \times \hat{\mathbf{B}}_{wave}$).

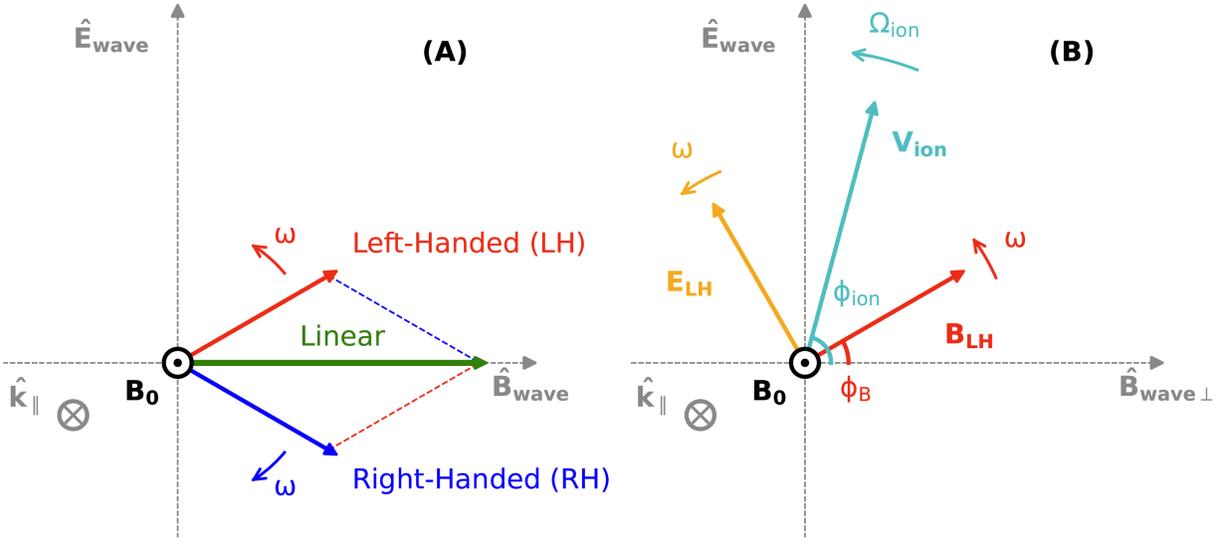

**Fig. S7. Visualization of the wave phase ($\phi_B$) and ion gyro phase ($\phi_{ion}$). (A)** The observed linear polarized waves (green line) can be regarded as a superposition of a pair of left-handed (red line) and right-handed (blue line) polarized waves with the same amplitude. **(B)** We define the wave phase ($\phi_B$) as the polar angle of the left-handed components of the magnetic field perturbation in the $\hat{B}_{wave} - \hat{E}_{wave}$ plane. Correspondingly, the polar angle of the ion velocity in the same plane is defined as the ion gyro phase ($\phi_{ion}$).

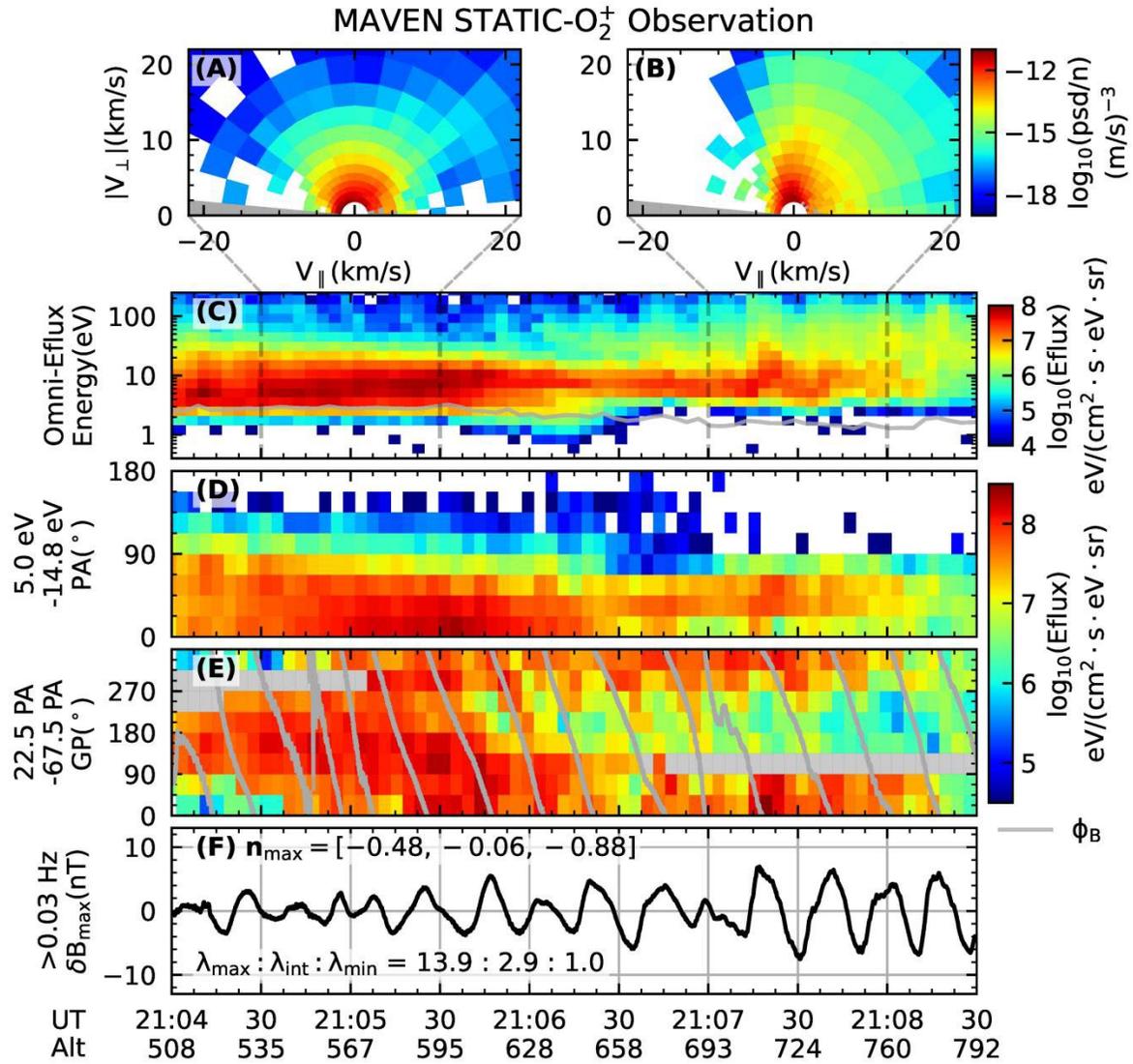

**Fig. S8. Kinetic features of $O_2^+$.** The format is the same as Fig. 3. The escape of $O_2^+$ is validated from its asymmetric velocity distribution function. Furthermore, an incomplete gyro phase stripes is observable in this result, indicating a non-resonant response to the waves of the $O_2^+$ [2].

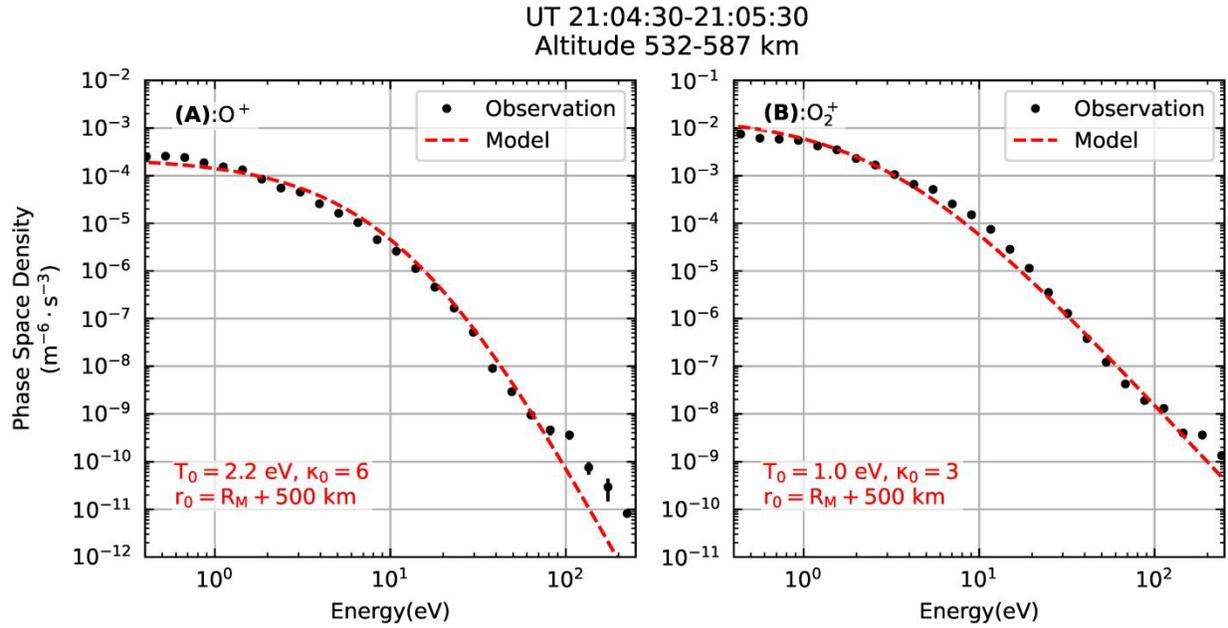

**Fig. S9. Comparison between the observed and modeled $O^+/O_2^+$ energy spectrum.** Considering the influence on the model spectrum from the gravitational potential energy, a Kappa distribution with gravitation correction is utilized here [3]. The spectrum model function is available at the supplementary material method (Kappa Distribution in the gravitational field) and the parameter is shown in the lower-left corner of each panel. Black error bars represent the counting uncertainty.

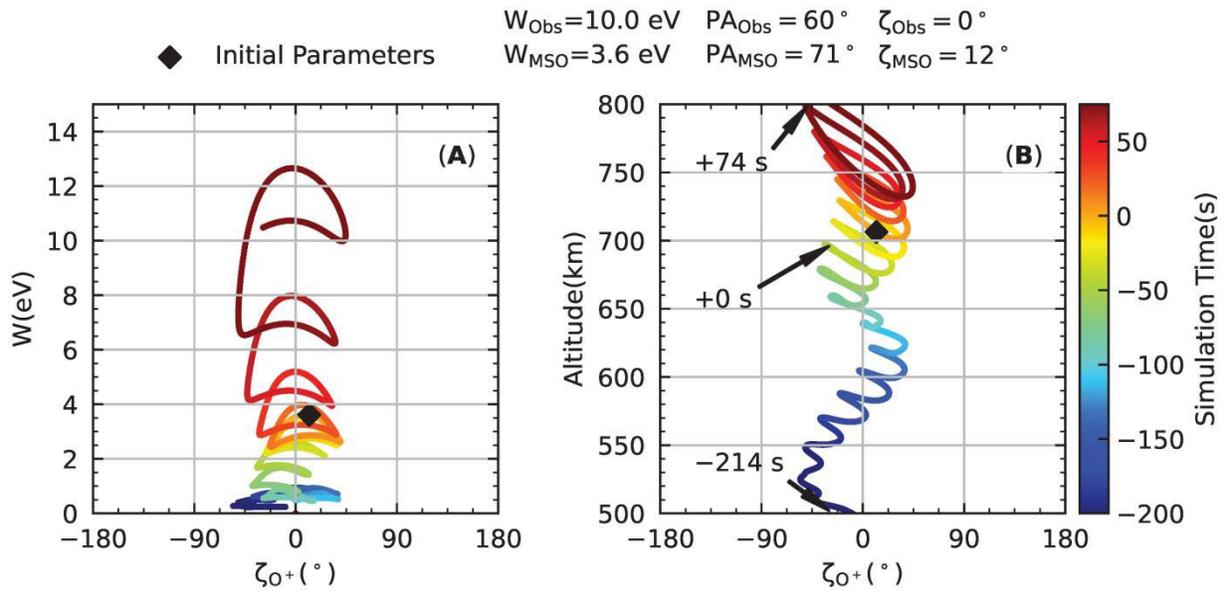

**Fig. S10. Characteristic ion trajectories in phase space.** (**A**) Simulated O$^+$ trajectory in the Energy (W)-$\zeta_{O+}$ space. (**B**) Simulated O$^+$ trajectory in the Altitude-$\zeta_{O+}$ space. The initial energy, pitch angle, and gyro phase of in the MSO coordinates and spacecraft coordinates (after spacecraft potential correction) are shown in the title. The test O$^+$ ion correspond to a phase bunched ion that observed by the spacecraft at UT 21:07:19 (as shown in Fig. 3**F**). Both backward and forward simulations are implemented to examine the ion behavior during its lifetime.

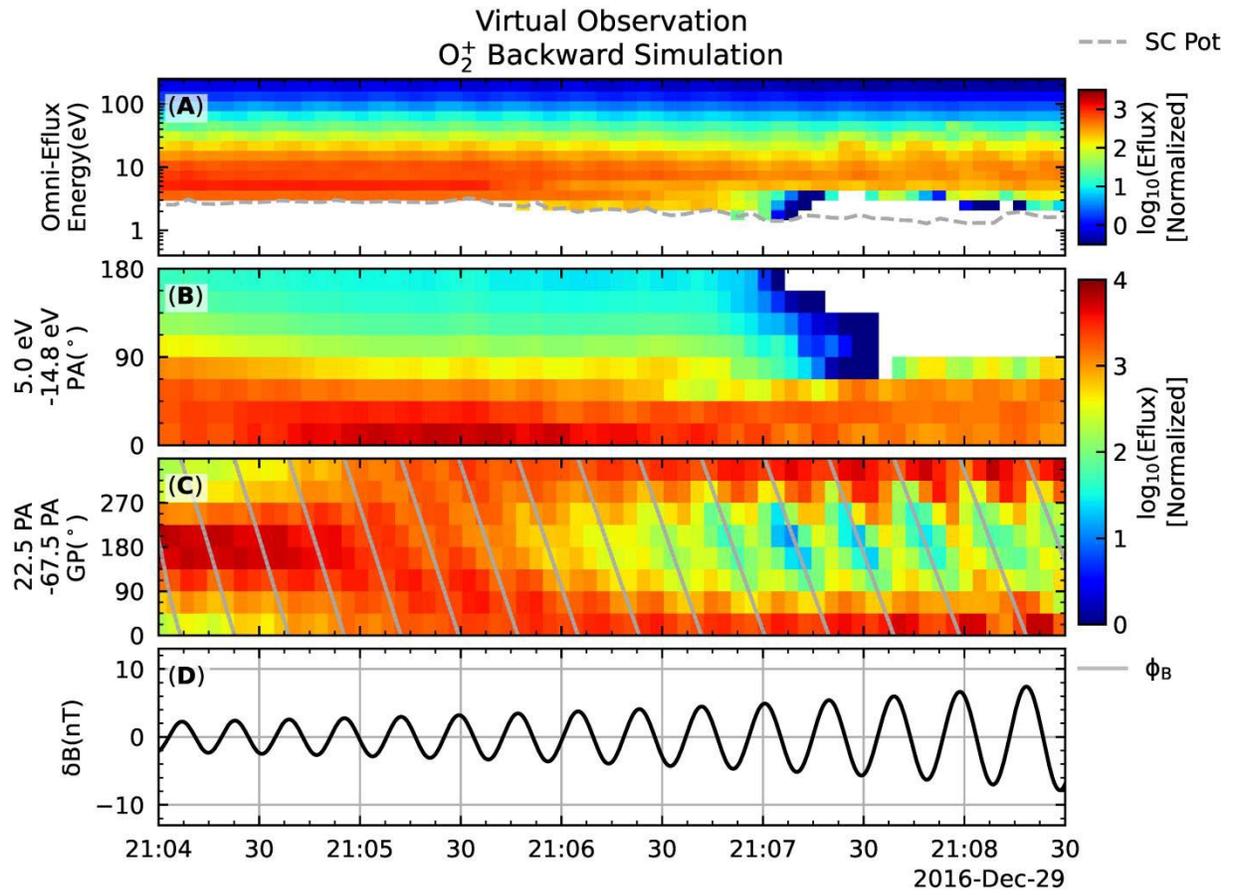

**Fig. S11. Virtual observation of backward Liouville simulation for $O_2^+$.** The format is the same as Fig. 5.

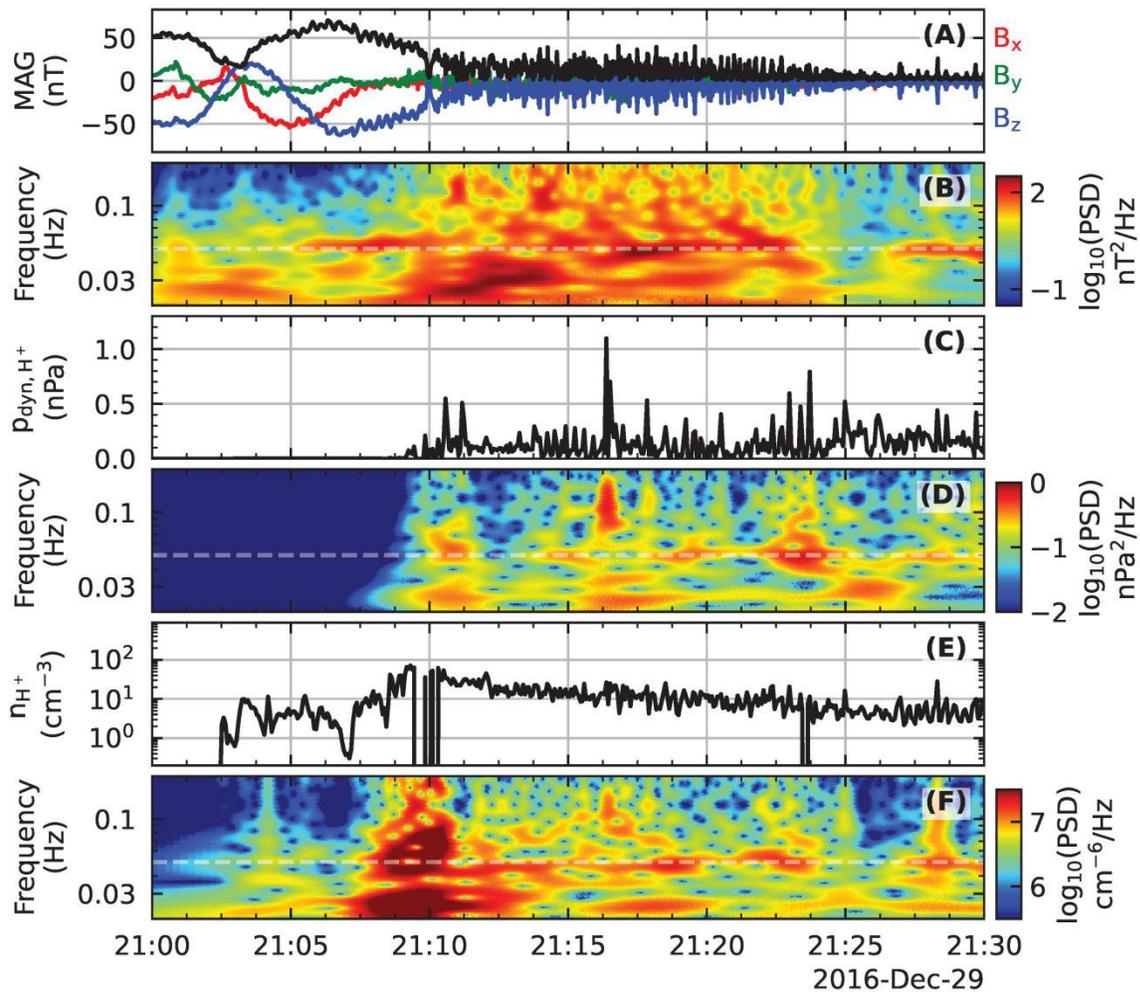

**Fig. S12. Perturbation analyses on the magnetic field, $H^+$ dynamic pressure, and $H^+$ number density. (A, B)** same as Figs. 2**D, E. (C, D)** Line plot and wavelet spectrum of $H^+$ dynamic pressure. **(E, F)** Line plot and wavelet spectrum of $H^+$ number density. Vertical dashed white lines at 0.05 Hz in the spectrum plots are used for guiding eyes. $H^+$ moments are obtained from the SWIA (Solar Wind Ion Analyzer) measurement [4]. [The wavelet power density shown in panels B, D, and F should be multiplied by a constant normalization factor. we find it after the submission, and we will revise it during the revision.]